\documentclass[preprint,12pt]{elsarticle}
\usepackage{amsmath}
\usepackage{color}

\usepackage{amssymb}

\usepackage{bm}
\usepackage{mathptmx} 

\journal{International Journal of Mechanical Sciences}

\begin{document}

\begin{frontmatter}



\title{Buckling patterns of complete spherical shells filled with 
an elastic medium under external pressure}


\author[huce,icl]{M.Sato\corref{cor1}}
\ead{tayu@eng.hokudai.ac.jp,m.sato@imperial.ac.uk}

\author[icl]{M.A.Wadee}
\ead{a.wadee@imperial.ac.uk}

\author[huce]{K.Iiboshi}
\ead{iiboshi@eng.hokudai.ac.jp}

\author[huce]{T.Sekizawa}
\ead{fsk-sfm-ttk@ec.hokudai.ac.jp}

\author[huap]{H.Shima}
\ead{shima@eng.hokudai.ac.jp}

\cortext[cor1]{Corresponding author}
\address[huce]{Division of Engineering and Policy for Sustainable
  Environment,  Faculty of Engineering, Hokkaido University, Kita 13
  Nishi 8, Kita-ku, Sapporo, Japan} 
\address[icl]{Department of Civil and Environmental Engineering,
  Imperial College London, SW7 2AZ, UK} 
\address[huap]{Division of Applied Physics, Faculty of Engineering,
  Hokkaido University, Kita 13 Nishi 8, Kita-ku, Sapporo, Japan} 

\begin{abstract}
  The critical buckling characteristics of hydrostatically pressurized
  complete spherical shells filled with an elastic medium are
  demonstrated.  A model based on small deflection thin shell theory,
  the equations of which are solved in conjunction with variational
  principles, is presented.  In the exact formulation, axisymmetric
  and inextensional assumptions are not used initially and the elastic
  medium is modelled as a Winkler foundation, \emph{i.e.}\ using
  uncoupled radial springs with a constant foundation modulus that is
  independent of wave numbers of shell buckling modes. Simplified
  approximations based on a Rayleigh--Ritz approach are also introduced
  for critical buckling pressure and mode number with a considerable
  degree of accuracy.  Characteristic modal shapes are demonstrated
  for a wide range of material and geometric parameters.  A phase
  diagram is established to obtain the requisite thickness to radius,
  and stiffness ratios for a desired mode profile.  The present exact
  formulation can be readily extended to apply to more general cases
  of non-axisymmetric buckling problems.
\end{abstract}


\end{frontmatter}



\section{Introduction}

The analytical study of the structural behaviour of spherical shells
is of great importance in the fields of not only civil, mechanical and
aeronautical engineering but also nanoscience and biomechanics;
Notable examples include pressure vessels, spherical honeycombs
\cite{TarnaiCMA1989}, carbon onions \cite{BaowanEPJD2007} and
spherical viruses \cite{RuJAP2009} and so on.  In the engineering
research fields, some pioneering theoretical works on the elastic
instability issues of empty complete spherical shells were conducted
in 1960s by Thompson \cite{ThompsonAQ1962}, Hutchinson
\cite{HutchinsonJAM1967} and Koiter \cite{KoiterPRDASB1969}.

In recent years, analytical works which considered an interaction
between a spherical shell and an internal elastic medium have also
been conducted.  The interaction effect leads to novel buckling
patterns which depend on the stiffness and thickness to radius ratios.
Recently for example, Yin \emph{et al.}\ \cite{YinPNAS2008}
successfully demonstrated the stress-driven buckling patterns in
spheroidal core/shell structures, where the core implies an elastic
medium, by using the finite element approach.  Interestingly, the
authors insisted in that work \cite{YinPNAS2008} that shapes of many
natural fruits and vegetables can be reproduced by anisotoropic
stress-driven buckles on the spheroidal core/shell system.  Moreover,
the morphology of a pressurized spherical shell can 
give clues to the structure--property relationship of spherical-shaped
nanostructures such as carbon onions \cite{PudlakPRA2009,LosPRB2009},
core/shell semiconductor nanoparticles \cite{TralleroPRB2010}, and the
``nano-matryoshka" \cite{ProdanScience2003}.  The last system refers
to a concentric multilayered structure comprising various metallic and
dielectric materials, showing great tunability in the optical
response; it is thus of interest to explore how pressure-induced
deformation affects the electromagnetic properties of the nanoshells
and their resonant frequencies.  

Purely from a structural mechanics perspective, Timoshenko
\cite{Timoshenko1973}, and Fl\"{u}gge \cite{Flugge1973} introduced the
formulation for buckling of a complete spherical shell in their
respective books that are now regarded as classics in the field.  In
these books, the exact approaches used to solve the hydrostatically
pressurized buckling of a complete spherical shell without an internal
elastic medium were introduced.  However, both of these formulations
were based on the axisymmetric assumption and formulations for more
general cases including non-axisymmetic deformations were not
included.  More recently, Fok and Allwrite \cite{FokJSA2001} analysed
the elastic axisymmetric buckling behaviour of a complete spherical
shell embedded in an elastic material and loaded by a far-field
hydrostatic pressure. In that study, the energy method in conjunction
with a Rayleigh--Ritz trial function was used for simplicity but the
validation of the obtained results was not discussed in detail with
buckling deformation modes being omitted.  Hence, to our knowledge, no
general non-axisymmetric formulation, in conjunction with exact
methods of solution, has been developed for the buckling behaviour of
a complete spherical shell with an internal elastic medium thus far.

Moreover, much attention has been given to the structural morphology
of the core/shell structure.  Some authors have recently demonstrated
the cross-sectional morphology of carbon nanotubes embedded in an
elastic medium \cite{ShimaNTN2008,ShimaPRB2010,SatoIJMPB2010}.
These results clearly show that interactions between shells and cores lead
to some novel wavy-shaped buckling deformation patterns.
 
From the background described above, the buckling properties of
hydrostatically pressurized complete spherical shells filled with an
elastic medium is demonstrated currently.  The elastic medium is
modelled as a Winkler foundation, \emph{i.e.}\ with uncoupled radial
springs and a constant foundation modulus.  An exact approach for
solving the developed equations based on the formulation without using
the axisymmetric and inextentional assumptions is presented.  This
approach therefore avoids any discussion about the validity of the
solution and allows to model to be extended to cover more generic
non-axisymmetric cases with relative ease.  The analytical results are
presented using a phase diagram and illustrative buckling modes.  In
addition to this, simplified approximations based on a Rayleigh--Ritz
approach are also introduced for the critical buckling pressure and
the corresponding mode number.  Comparative studies between the exact
and simplified approaches are conducted to validate the approximation
and it is shown that the approximate formulations enables sufficiently
accurate values to be obtained.

\section{Complete spherical shell model}

\begin{figure}
\centering
\includegraphics[width=8cm]{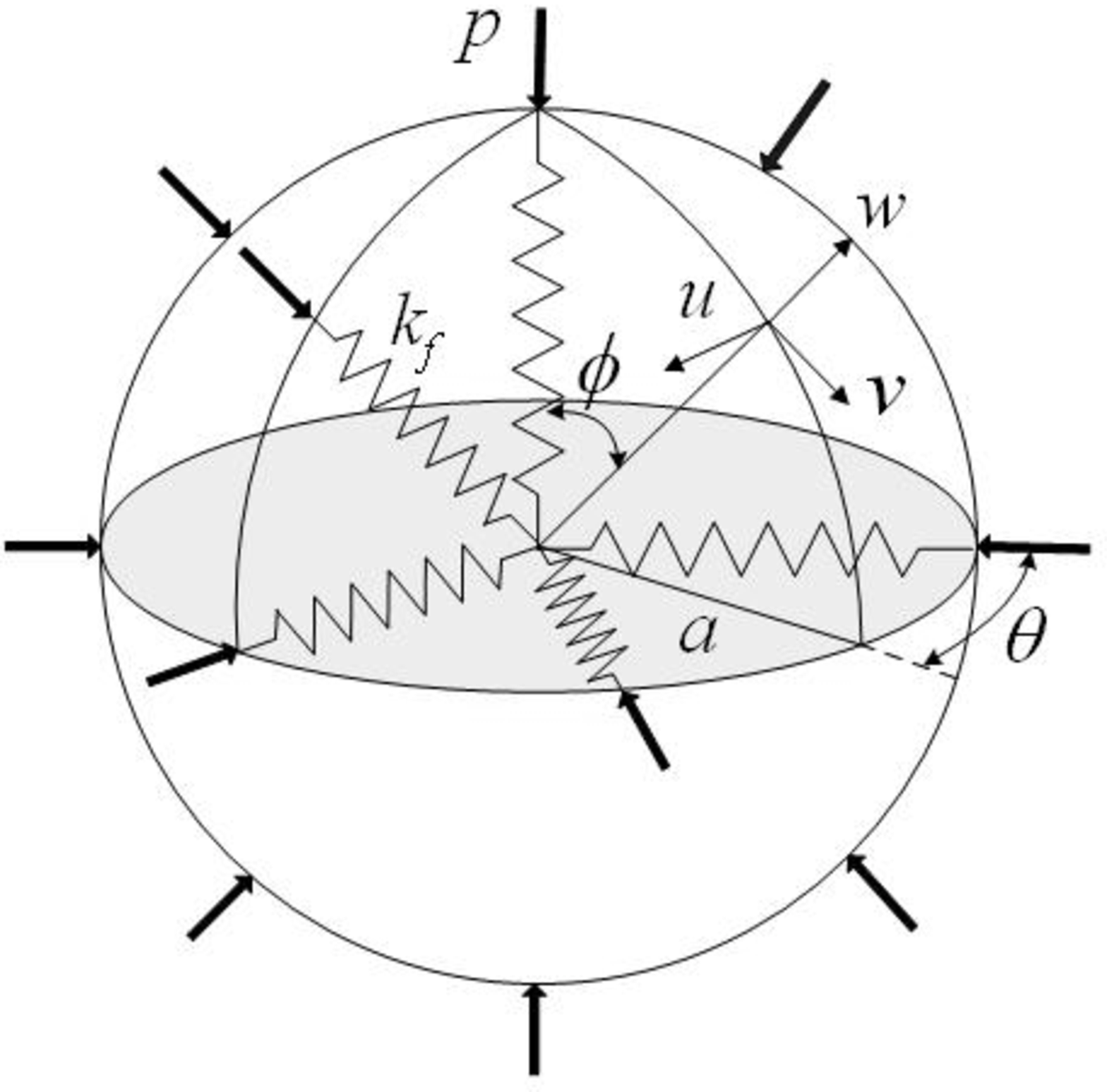}
\caption{Hydrostatically pressurized spherical shell filled with an
  elastic medium.}
\label{AnalyticalModel}
\end{figure}

The critical pitchfork bifurcation phenomenon of the hydrostatically
pressurized complete spherical shell is investigated, as shown in
Fig.\ref{AnalyticalModel}.  The spherical shell is constructed from an
homogeneous and isotropic linear elastic material with Young's modulus
$E$ and Poisson's ratio $\nu$, which is filled with an elastic
material that is modelled as a Winkler foundation, \emph{i.e.}\ with
uncoupled springs in the radial direction and a constant foundation
modulus $k_f$.  For a complete spherical shell with radius $a$ and
thickness $h$, spherical angular coordinates in the latitude and
meridian directions $\theta$ and $\phi$ are used. Displacement
functions: $u$, $v$ and $w$ are in the $\theta$, $\phi$ and the
(outward) radial directions, respectively.

\section{Formulation}

\subsection{Exact approach}

\subsubsection{Energy formulation}

The following analysis is based on classical small deformation theory
of thin shells~\cite{Brush1975,CrollJSV1980}.  The total potential
energy $V$ is expressed by the sum of the strain energy and work
done by external load as:
\begin{equation}
V = U_M + U_B + U_F + \Omega,
\label{TPE}
\end{equation}
in which $U_M$ is the membrane (in-plane) strain energy term, where:
\begin{eqnarray}
  && U_M = U_{M\phi} + U_{M\theta} + U_{M\phi\theta},  \nonumber \\ 
  && U_{M\phi} = \frac{1}{2}\int_\phi \int_\theta N_\phi \epsilon_\phi
  a^2 \sin \phi~\mathrm d \phi \mathrm d \theta,  \nonumber \\
  && U_{M\theta} = \frac{1}{2}\int_\phi \int_\theta N_\theta
  \epsilon_\theta a^2 \sin \phi~\mathrm d \phi~\mathrm d \theta,  \nonumber \\
  && U_{M\phi\theta} = \int_\phi \int_\theta N_{\phi\theta}
  \epsilon_{\phi\theta} a^2 \sin \phi~\mathrm d \phi~\mathrm d \theta,
\label{U_membrane}
\end{eqnarray}
$U_B$ is the bending (out-of-plane) component, where:
\begin{eqnarray}
&& U_B = U_{B\phi} + U_{B\theta} + U_{B\phi\theta},  \nonumber \\
&& U_{B\phi} = \frac{1}{2}\int_\phi \int_\theta M_\phi \chi_\phi a^2
\sin \phi~\mathrm d \phi~\mathrm d \theta,  \nonumber \\
&& U_{B\theta} = \frac{1}{2}\int_\phi \int_\theta M_\theta \chi_\theta
a^2 \sin \phi~\mathrm d \phi~\mathrm d \theta,  \nonumber \\
&& U_{B\phi\theta} = \int_\phi \int_\theta M_{\phi\theta}
\chi_{\phi\theta} a^2 \sin \phi~\mathrm d \phi~\mathrm d \theta,
\label{U_bending}
\end{eqnarray}
$U_F$ is the strain energy term due to a Winkler foundation, thus:
\begin{equation}
U_{F} = \frac{1}{2}\int_\phi \int_\theta a^2 k_f w^2 \sin
\phi~\mathrm d \phi~\mathrm d \theta,
\label{U_foundation}
\end{equation}
and $\Omega$ is the potential energy of the applied pressure:
\begin{equation}
\Omega = \int_\phi \int_\theta a^2 p w \sin \phi~\mathrm d \phi~\mathrm d \theta.
\label{Omega}
\end{equation}

Based on the assumptions of thin shell theory \cite{Brush1975}, the
strain--displacement relations can be expressed as:
%
\begin{eqnarray}
&& \epsilon_\phi = \frac{1}{a}(v,_{\phi}+w),     \nonumber \\
&& \epsilon_\theta = \frac{1}{a}
\left(
\cot\phi v + \frac{u,_{\theta}}{\sin\phi} + w
\right),   \nonumber \\
&& \epsilon_{\phi\theta} = \frac{1}{a}
\left(
\frac{v,_{\theta}}{\sin\phi} -
\cot\phi u + u,_{\phi}
\right),  \nonumber \\ 
&& \beta_\phi = \frac{1}{a}(-w,_{\phi} + v),     \nonumber \\
&& \beta_\theta = \frac{1}{a}
\left(
-\frac{w,_{\theta}}{\sin\phi} + u
\right),
\nonumber \\
&& \chi_\phi = \frac{1}{a^2}(-w,_{\phi\phi} + v,_{\phi}),     \nonumber \\
&& \chi_\theta = \frac{1}{a^2}
\left[
-\frac{w,_{\theta\theta}}{\sin^{2}\phi}
+ u,_{\theta} + \cot\phi(v-w,_\phi)
\right], \nonumber \\ 
&& \chi_{\phi\theta} = \frac{\sin\phi u,_\phi + v,_\theta
  -2w,_{\phi\theta} + 2\cot\phi w,_\theta - \cos\phi u} {2a^2\sin\phi},
\nonumber \\
\label{strain_displacement}
\end{eqnarray}
%
where ``$,x$'' denotes differention with respect to $x$, and the
constitutive relations are as follows:
\begin{eqnarray}
&& N_\phi = C (\epsilon_\phi + \nu \epsilon_\theta), \ \  N_\theta = C
(\epsilon_\theta + \nu \epsilon_\phi),   \nonumber \\
&& N_{\phi\theta} = C \frac{1-\nu}{2} \epsilon_{\phi\theta}, \ \
M_\phi = D (\chi_\phi + \nu \chi_\theta),  \nonumber \\
&& M_\theta = D (\chi_\theta + \nu \chi_\phi), \ \  M_{\phi\theta} = D
\frac{1-\nu}{2} \chi_{\phi\theta},
\label{constitutive_relation}   
\end{eqnarray}
where $C = Eh/(1-\nu^2)$ is the membrane stiffness and $D =
Eh^3/12(1-\nu^2)$ is the flexural rigidity of the spherical shell,
respectively.  Substituting the expressions in
Eqs.(\ref{strain_displacement}) and (\ref{constitutive_relation}) into
Eqs.(\ref{U_membrane})--(\ref{U_foundation}) gives the energy
expressions in terms of the displacements $u$, $v$ and $w$.

\subsubsection{Fundamental state}

When uniform hydrostatic pressure $p$ is acting on the spherical
shell, only the inward radial static displacement $w_0$ can occur in
the pre-buckling state.  Hence, the stress-strain relationship of the
spherical shell is assumed to be linear up to the point of
instability.  In the fundamental pre-buckling state, the potential
energies $V^{(0)}$ are given by:
\begin{equation}
V^{(0)} = U_M^{(0)} + U_B^{(0)} + U_F^{(0)} + \Omega^{(0)} =
\int_\phi \int_\theta F^{(0)}~\mathrm d \phi~\mathrm d\theta 
\label{U0}
\end{equation}
where
\begin{eqnarray}
&& U_M^{(0)} = C (1+\nu) \int_\phi \int_\theta w_0^2 ~\mathrm d \phi
~\mathrm d \theta, \ \  U_B^{(0)} = 0,  \nonumber \\
&& U_F^{(0)} = \frac{1}{2}\int_\phi \int_\theta a^2 k_f w^2 \sin \phi
~\mathrm d \phi ~\mathrm d \theta, \nonumber \\
&& \Omega^{(0)} = \int_\phi \int_\theta a^2 p w_0 \sin \phi ~\mathrm d \phi
~\mathrm d \theta.
\label{Ustatic}
\end{eqnarray}
For an equilibrium state, the first variation of the total potential
energy $V$ must equal zero.  This condition gives the following static
displacement under uniform hydrostatic pressure:
\begin{equation}
w_0 = - \frac{a^2 (1-\nu)}{2Eh + a^2 k_{f} (1-\nu)} p.
\label{w0}
\end{equation}

\subsubsection{Critical buckling analysis}

To obtain an expression for the second variation of the total
potential energy, the following infinitesimally small increments are
defined:
\begin{equation}
u = u_1, \ \  v = v_1, \ \ w = w_0 + w_1,
\label{uvw}   
\end{equation}
which correspond to buckling displacement modes.  The expression for
the potential energy due to the external force is a linear functional
of the displacement components and makes no contribution to the second
variation expression, which is $\delta^2\Omega=0$ \cite{Brush1975}.
Consequently, the second variation of the total potential energy
becomes
\begin{equation}
\delta^2 U = \delta^2 U_M + \delta^2 U_B + \delta^2 U_F = \int_\phi
\int_\theta F ~\mathrm d \phi ~\mathrm d \theta,
\label{delta2U}   
\end{equation}  
where
\begin{align}
  \delta^2 U_{M} =&\delta^2 U_{stretch} + \delta^2 U_{shear},  \nonumber \\
  \delta^2 U_{stretch} =&\frac{C w_0 (1+\nu)}{a} \int_\phi
  \int_\theta [( v_1 + w_1,_\phi )^2    
   + ( u_1 - w_1,_\theta \csc\phi)^2 ] \sin\phi ~\mathrm d \phi
  ~\mathrm d \theta   \nonumber \\ 
  & + C \int_\phi \int_\theta [( v_1 + w_1,_\phi )^2
   + ( u_1 - w_1,_\theta \csc\phi)^2 ]  ~\mathrm d \phi ~\mathrm d \theta,
  \nonumber \\
  \delta^2 U_{shear} =&\frac{C(1-\nu)}{2} \int_\phi
  \int_\theta (\cot\phi u_1 + \csc\phi v_1,_\theta  
   + u_1,_\phi )^2 \sin\phi ~\mathrm d \phi ~\mathrm d \theta,
\label{delta2UM}   
\end{align}
\begin{align}
  \delta^2 U_{B} = & \delta^2 U_{bend} + \delta^2 U_{twist}, \nonumber \\
  \delta^2 U_{bend} = & \frac{D}{a^2} \int_\phi \int_\theta \{(\cot\phi
  u_1 + \csc\phi v_1,_\theta   
  -\csc^2\phi w_1,_{\theta\theta} - \cot\phi w_1,_\phi )^2    \nonumber \\
  & (v_1,_\phi + w_1,_{\phi\phi})[v_1,_\phi + w_1,_{\phi\phi}+  
   2\nu (\cot\phi v_1 + \csc\phi u_1,_\theta - \csc^2\phi
  w_1,_{\theta\theta}   \nonumber \\
  & - \cot\phi w_1,_\phi)] \} \sin\phi ~\mathrm d \phi ~\mathrm d \theta , \nonumber \\
  \delta^2 U_{twist} = & \frac{D (1-\nu)}{2a^2} \int_\phi \int_\theta
  [2 (1+\nu) w_1 
   ( v_1 \cot\phi + w_1 + u_1,_\theta + u_1,_\phi)   \nonumber \\
  & + u_1,_\phi (2\nu v_1 \cot\phi + 2\nu u_1,_\theta + v_1,_\phi)  \nonumber \\
  & +(v_1 \cot\phi + u_1,_\theta \csc\phi)^2] \sin\phi ~\mathrm d \phi
  ~\mathrm d \theta, \nonumber \\
  \label{delta2UMphitheta}   
\end{align}
\begin{equation}
\delta^2 U_F = \int_\phi \int_\theta a^2 k_F w_1^2 \sin \phi d\phi d\theta.
\label{delta2UF}   
\end{equation}

According to the Trefftz criterion, the buckling equations can be
obtained by introducing $F$ in Eq.(\ref{delta2U}) into the
Euler--Lagrange equations with the calculus of variations.
The Euler--Lagrange equations in this case are as
follows \cite{Brush1975}:
\begin{align}
  & \frac{\partial F}{\partial u_1} - \frac{\partial}{\partial \phi}
  \left(\frac{\partial F}{\partial u_{1,\phi}}\right) 
  - \frac{\partial}{\partial \theta} \left( \frac{\partial F}{\partial
      u_{1,\theta}} \right) = 0, \nonumber \\
  & \frac{\partial F}{\partial v_1} - \frac{\partial}{\partial
    \phi} \left( \frac{\partial F}{\partial v_{1,\phi}} \right) 
  - \frac{\partial}{\partial \theta} \left( \frac{\partial F}{\partial
      v_{1,\theta}} \right) = 0, \nonumber \\ 
  & \frac{\partial F}{\partial w_1} - \frac{\partial}{\partial \phi}
  \left( \frac{\partial F}{\partial w_{1,\phi}} \right) -
  \frac{\partial}{\partial \theta} \left( \frac{\partial F}{\partial
      w_{1,\theta}} \right)  \nonumber \\
  & \quad + \frac{\partial^2}{\partial \phi^2} \left(\frac{\partial^2
      F}{\partial w_{1,\phi\phi}} \right)   
  + \frac{\partial^2}{\partial \phi \partial \theta} \left(
    \frac{\partial^2 F}{\partial w_{1,\phi\theta}} \right) 
  + \frac{\partial^2}{\partial \theta^2} \left( \frac{\partial^2
      F}{\partial w_{1,\theta\theta}} \right)  = 0.
\label{Euler}   
\end{align}
 
Substitution of a solution of the form
\begin{eqnarray}
u_1 = \sum \bar{u}(\phi) \sin n \theta, \nonumber \\
v_1 = \sum \bar{v}(\phi) \cos n \theta, \nonumber \\
w_1 = \sum \bar{w}(\phi) \cos n \theta,
\label{u1v1w1}   
\end{eqnarray}
where $n$ is a positive integer, into
Eqs.(\ref{delta2U})--(\ref{delta2UF}) and some rearrangement gives the
following ordinary differential equations for an arbitrary $n$
\begin{equation}
a_u^{(i)} \frac{\partial^i \bar{u}}{\partial \phi^i} + b_u^{(i)}
\frac{\partial^i \bar{v}}{\partial \phi^i} 
+ c_u^{(i)} \frac{\partial^i \bar{w}}{\partial \phi^i}   
- \frac{w_0}{a}(1+\nu) (\bar{u} \sin\phi + n \bar{w}) = 0,
\label{ODEu-direction}   
\end{equation}
\begin{equation}
  a_v^{(i)} \frac{\partial^i \bar{u}}{\partial \phi^i} + b_v^{(i)}
  \frac{\partial^i \bar{v}}{\partial \phi^i} 
  + c_v^{(i)} \frac{\partial^i \bar{w}}{\partial \phi^i}   
  - \frac{w_0}{a}(1+\nu) \sin\phi ( \bar{v} - \frac{\partial
    \bar{w}}{\partial \phi}) = 0, 
\label{ODEv-direction}   
\end{equation}
\begin{align}
  & a_w^{(i)} \frac{\partial^i \bar{u}}{\partial \phi^i} + b_w^{(i)}
  \frac{\partial^i \bar{v}}{\partial \phi^i} 
  + c_w^{(i)} \frac{\partial^i \bar{w}}{\partial \phi^i} +
  \frac{a^2 k_f}{C} \bar{w}
  \sin\phi  \nonumber \\ 
  & \quad + \frac{w_0}{a}( 1+\nu ) \sin\phi \Bigl( \frac{\partial
    \bar{v}}{\partial \phi} + \bar{v} \cot\phi +
  \frac{n\bar{u}}{\sin\phi}   
  - \frac{\partial^2 \bar{w}}{\partial \phi^2} -
  \frac{\partial \bar{w}}{\partial \phi}\cot\phi + \frac{n^2}{\sin^2
    \phi} \bar{w} \Bigr ) = 0,
\label{ODEw-direction}   
\end{align}
in which the implied summation over $i$ cover the range of non-zero
coefficients, which are described in the Appendix.
These governing differential equations
(\ref{ODEu-direction})--(\ref{ODEw-direction}) are difficult to solve
exactly by only assuming periodic functions in $\bar{u}$, $\bar{v}$
and $\bar{w}$ directly.  However, the definition of the following
special linear differential operator enables the solution of these
equations with an exact approach:
\begin{equation}
  H(...) = (...)'' + (...)'\cot\phi + (...) \left( 2 -
    \frac{n^2}{\sin^2\phi} \right), 
\label{H}   
\end{equation}
where primes denotes differentiation with respect to $\phi$.  From the
definition of $H$ in Eq.(\ref{H}), the following relations are readily
obtained:
\begin{align}
  HH(...) = & (...)'''' + 2(...)'''\cot\phi 
  + (...)''\left( 3 - \frac{1 + 2 n^2}{\sin^2\phi} \right) \nonumber \\
  & + (...)'\cot\phi \left( 4 + \frac{1 + 2 n^2}{\sin^2\phi} \right)
  + (...) \left( 4 - \frac{2 n^2}{\sin^2\phi} - \frac{n^2 (4 -
      n^2)}{\sin^4\phi} \right),
  \label{HH}   
\end{align}
\begin{equation}
  H'(...) = (...)''' + (...)''\cot\phi  
  + (...)'\cot\phi \left( 2 - \frac{1 +  n^2}{\sin^2\phi} \right)
  + 2n^2 (...) \frac{\cos\phi}{\sin^3\phi}.   
  \label{Hdash}   
\end{equation}
In addition, a new variable $\hat{u}$ is defined with regard to
$\bar{u}$ by
\begin{equation}
\bar{u} = \hat{u}'.
\label{U}  
\end{equation}
By making use of the relations of Eqs.(\ref{H})--(\ref{Hdash}) and the
new definition of Eq.(\ref{U}), Eqs.(\ref{ODEu-direction}) and
(\ref{ODEv-direction}) can be rewritten respectively as:
\begin{align}
  & (1+k) \sin\phi \frac{\partial}{\partial\phi} \left(
    \frac{1-\nu}{2} H(\hat{u}) - \frac{1+\nu}{2}n \frac{\bar{v}
      \sin\phi + n\hat{u}}{\sin^2 \phi} \right)
  - 2(1+k)n \frac{\bar{v} \sin\phi + n\hat{u}}{\sin^2 \phi}\cos\phi
  \nonumber \\
  & \quad -(1+k)(1+\nu)nw + knH(w) -\frac{w_0 (1+\nu)}{a} (\hat{u}' +
  n\bar{w}) = 0,
\label{Hu}  
\end{align}
\begin{align}
& (1+k) \biggl[\frac{1+\nu}{2}(\bar{v} \sin\phi + n\hat{u})'' +
 \frac{1-\nu}{2} H(\bar{v} \sin\phi)  
 - \frac{3-\nu}{2}\cot\phi(\bar{v} \sin\phi + n\hat{u})' + \nonumber \\
& \quad (1+\nu)w'\sin\phi \biggr]    
 - k\sin\phi H'(w) -\frac{w_0 (1+\nu)}{a}\sin\phi (\bar{v} -
\bar{w}')= 0.
\label{Hv}  
\end{align}
Differentiation of Eq.(\ref{Hv}) with respect to $\phi$ and then
multiplying it by $\sin\phi/n$ leads to the following:
\begin{align}
  & (1+k) \frac{\sin\phi}{n} \biggl\{\frac{1+\nu}{2}(\bar{v} \sin\phi +
  n\hat{u})''' + \frac{1-\nu}{2} H'(\bar{v} \sin\phi)  \nonumber \\
  & + \frac{3-\nu}{2} \biggl[\frac{1}{\sin^2\phi}(\bar{v} \sin\phi + n\hat{u})'
  -\cot\phi(\bar{v} \sin\phi + n\hat{u})'' \biggr]   \nonumber \\
  & + (1+\nu)(w''\sin\phi + w'\cos\phi) \biggr\}   
  - k\frac{\sin\phi}{n} [\cos\phi H'(w) +\sin\phi H''(w) ]= 0. 
\label{Hvd}  
\end{align}
By adding Eq.(\ref{Hvd}) to Eq.(\ref{Hu}), some rearrangement gives:
\begin{eqnarray}
  && (1+k) \left[H(\Psi) - (1+\nu)(\Psi -H(w) +2w) \right]   \nonumber \\
  && \quad -k \left( HH(w)-2H(w) \right) -\frac{w_0}{a}(1+\nu)(\Psi-H(w)+2w) =
  0. \nonumber \\
\label{Huv}  
\end{eqnarray}
in which:
\begin{equation}
\Psi = \frac{(\bar{v} \sin\phi + n\hat{u})'}{\sin\phi}.
\label{Psi}  
\end{equation}

Similarly, Eq.(\ref{ODEw-direction}) can be written using the operator
$H$ and the new variable $\Psi$ defined by Eq.(\ref{Psi}) as follows:
\begin{align}
  & (1+k)(1+\nu)(\Psi+2w)  
   +k[HH(w)-(3+\nu)H(w)-H(\Psi)]    \nonumber \\
  & \quad +\frac{a^2 k_F}{C}w +(1+\nu) \frac{w_0}{a}(\Psi-H(w)+2w) = 0. 
\label{Hw}  
\end{align}

Equations (\ref{Huv}) and (\ref{Hw}) are the governing equations to be
solved.  Now, the solutions of $\Psi$ and $w$ of Eq.(\ref{Huv}) and
(\ref{Hw}) is assumed to be in the form
\begin{equation}
  \Psi = \sum_{m=0}^\infty A_m P_m (\cos\phi),\ \ w =
  \sum_{m=0}^\infty B_m P_m (\cos\phi),
\label{Psiw}   
\end{equation}
where $A_m$ and $B_m$ are the constant deformation amplitudes, and the
spherical harmonic $P_m(\cos\phi)$ is a series of Legendre functions
of degree $n$ which satisfies the following Legendre differential
equation,
\begin{equation}
  P_m'' + P_m'\cot\phi + m(m+1)P_m = 0.
\label{Pm}   
\end{equation}
By using the operator $H$ defined in Eq.(\ref{H}), the following
relation can be obtained:
\begin{equation}
H(P_m) = -\lambda_m P_m, 
\label{HPm}   
\end{equation}
where
\begin{equation}
\lambda_m = m(m+1)-2,
\label{lambdam}
\end{equation}
and
\begin{equation}
HH(P_m) = -\lambda^2_m P_m.
\label{HHPm}   
\end{equation}
Substituting the series of Eq.(\ref{Psiw}) into Eqs.(\ref{Huv}) and
(\ref{Hw}) and using Eqs.(\ref{HPm}) and (\ref{HHPm}), we obtain:
\begin{eqnarray}
  \sum_{m=0}^\infty (c_{m}^{(11)} A_m + c_{m}^{(12)} B_m ) P_m
  (\cos\phi) = 0, \nonumber \\ 
  \sum_{m=0}^\infty (c_{m}^{(21)} A_m + c_{m}^{(22)} B_m ) P_m (\cos\phi) = 0,
\label{be}  
\end{eqnarray}
in which
\begin{align}
  c_{m}^{(11)} = & (1+k)(\lambda_m + 1+\nu) + \frac{w_0}{a}(1+\nu), \nonumber \\
  c_{m}^{(12)} = & (1+k)(1+\nu)(\lambda_m+2) + k(\lambda_m^2+2k\lambda_m) 
   + \frac{w_0}{a}(1+\nu)(1+\lambda_m), \nonumber \\
  c_{m}^{(21)} = & (1+k)(1+\nu) + k\lambda_m + \frac{w_0}{a}(1+\nu),
  \nonumber \\
  c_{m}^{(22)} = & 2(1+k)(1+\nu) + k\lambda_m(\lambda_m+3+\nu)  
   + \frac{a^2 k_F}{C} + \frac{w_0}{a}(1+\nu)(\lambda_m+2). 
\label{cmij}  
\end{align}

Hence, homogeneous linear equations for $A_m$ and $B_m$ are obtained
for all $m$ as follows:
\begin{equation}
\left[
\begin{array}{cc}
 c_{m}^{(11)} & c_{m}^{(12)} \\  c_{m}^{(21)} & c_{m}^{(22)} 
\end{array} 
\right]
\left(
\begin{array}{c}
 A_m \\ B_m 
\end{array} 
\right)
=\left(
\begin{array}{c}
 0 \\ 0 
\end{array} 
\right).
\label{det1}   
\end{equation}
This is a standard eigenvalue problem for which nontrivial values are
obtained when the coefficient matrix in Eq.(\ref{det1}) becomes
singular.  The corresponding pressure $p$ is thus calculated from the
determinantal equations:
\begin{equation}
\det \left[
\begin{array}{cc}
 c_{m}^{(11)} & c_{m}^{(12)} \\  c_{m}^{(21)} & c_{m}^{(22)} 
\end{array} 
\right] = 0.
\label{det2}
\end{equation}

By solving Eq.(\ref{det2}), the pressure $p$ with the corresponding
mode $m$ for several combinations of geometric and material constants
can be obtained as an eigenvalue.  For the particular values of the
constants, the minimum value of the eigenvalues is the critical
buckling pressure $p_{cr}$.  It should be noted that the above
equation is independent of the mode number $n$.  This shows that only
axisymmetric modes can occur in this problem despite the inclusion
of the Winkler foundation term.

\subsection{Simplified approach}

Next, we consider the simplified formulation by the Rayleigh--Ritz approach.
Here an inextensional and axisymmetric buckling deformation of the shell are assumed with 
shear and torsional strains being neglected for simplicity.
Thus, $u$, $\epsilon_{\phi\theta}$, and $\chi_{\phi\theta}$ are assumed to be zero.

The buckling deformation forms are given as follows:     
\begin{equation}
v_1 = \sqrt{1-c^2} \sum_{m=2}^{\infty} \tilde{v}_m \frac{dP_m(c)}{dc}, 
\label{v1}   
\end{equation}
\begin{equation}
w_1 = \sum_{m=2}^{\infty} \tilde{w}_m P_m(c), 
\label{w1}   
\end{equation}
where $c=\cos\phi$.

The first variation $\delta U_M$ of the membrane (in-plane) strain energies is given by
\begin{equation}
\delta U_M = 2\pi C (1+\nu) \int_\phi w_0 \left(v'_1 + v_1 \cot\phi + 2w_1 \right) \sin\phi d\phi.
\label{deltaUM-RR}   
\end{equation}
The integrand in Eq.(\ref{deltaUM-RR}),
\begin{equation}
v'_1 + v_1 \cot\phi + 2w_1 =0,
\label{inextensional}   
\end{equation}
gives the inextensional buckling condition and when the assumed
deformation functions in Eq.(\ref{v1}) and (\ref{w1}) satisfy the
condition, the following relation can be obtained
\begin{equation}
\tilde{w}_m = \frac{m(m+1)}{2} \tilde{v}_m.
\label{w-v}   
\end{equation}
The second variation of the strain energies are
\begin{equation}
\delta^2 V = \delta^2 U_M + \delta^2 U_B + \delta^2 U_F = 2\pi \int_\phi F d\phi,
\label{delta2U-RR}   
\end{equation}  
in which the each term can be expressed by the strain--displacement
relations and Eqs.(\ref{v1}) and (\ref{w1}) as:
\begin{eqnarray}
  \delta^2 U_M = \pi C \sum_{m=2}^{\infty} \Biggl
  [\frac{2(m-1)m(m+1)(m+2)(1-\nu)}{2m+1}   \nonumber \\
  +\frac{w_0}{a}\frac{(m-1)^2m(m+1)(m+2)^2(1+\nu)}{2m+1} \Biggr] \tilde{v}_m^2,
\label{delta2UM-RR}   
\end{eqnarray}
\begin{equation}
\delta^2 U_B = \frac{\pi D}{a^2} \sum_{m=2}^{\infty} \frac{(m-1)^2
  m(m+1)(m+2)^2[m(m+1)-1+\nu]}{2m+1} \tilde{v}_m^2, 
\label{delta2UB-RR}   
\end{equation}
\begin{equation}
\delta^2 U_F = \pi a^2 k_F \sum_{m=2}^{\infty} \frac{m^2(m+1)^2}{2m+1} \tilde{v}_m^2 .
\label{delta2UF-RR}   
\end{equation}
The stability criterion
\begin{equation}
\frac{\partial(\delta^2 V)}{\partial \tilde{v}_m} = 0,
\label{d(d2U)dVm}   
\end{equation}
gives the following critical pressure
\begin{equation}
\begin{split}
  p_{cr} = & \frac{2Eh + k_f a^2(1-\nu)}{aEh (m-1)^2 (m+2)^2} \Biggl
  \{\frac{2Eh + (m-1)(m+2)}{1+\nu}  \\
  & + \frac{D (m-1)^2(m+2)^2[m(m+1)-1+\nu]}{a^2} + a^2 k_f m(m+1)
  \Biggr \}.
\label{pcr-RR}
\end{split}   
\end{equation}
Equation (\ref{pcr-RR}) has a minimum value when
\begin{equation}
\frac{\partial p_{cr}}{\partial m} = 0.
\label{dpcrdm}   
\end{equation}
In such spherical shell buckling, $m$ is usually a large integer.
This allows us to approximate $(m-1)^2 (m+2)^2 \approx m^4$, $m(m+1)
\approx m^2$, and so on.  By applying the approximations to
Eq.(\ref{pcr-RR}) and substituting the resulting equation into
(\ref{dpcrdm}), the mode number $\tilde{m}_{cr}$ associated with the
lowest critical pressure and the corresponding critical pressure
$\tilde{p}_{cr}$ can be obtained as follows:
\begin{equation}
\tilde{m}_{cr} = \sqrt[4]{\frac{a^2 [2Eh +a^2 k (1+\nu)]}{D(1+\nu)}} , 
\label{ncr}   
\end{equation}
\begin{equation}
  \tilde{p}_{cr} = \frac{2Eh + ka^2(1-\nu)}{aEh \tilde{m}_{cr}^2}
  \left[\frac{2Eh + ka^2(1+\nu)}{1+\nu} + \frac{D \tilde{m}_{cr}^2
      (\tilde{m}_{cr}^2 -1 +\nu)}{a^2} \right].
\label{pcr-s}   
\end{equation}

\section{Analytical results and discussion}

\subsection{Comparison of buckling pressures between exact and
  simplified approaches}

\begin{figure}
\centering
\includegraphics[width=10.5cm]{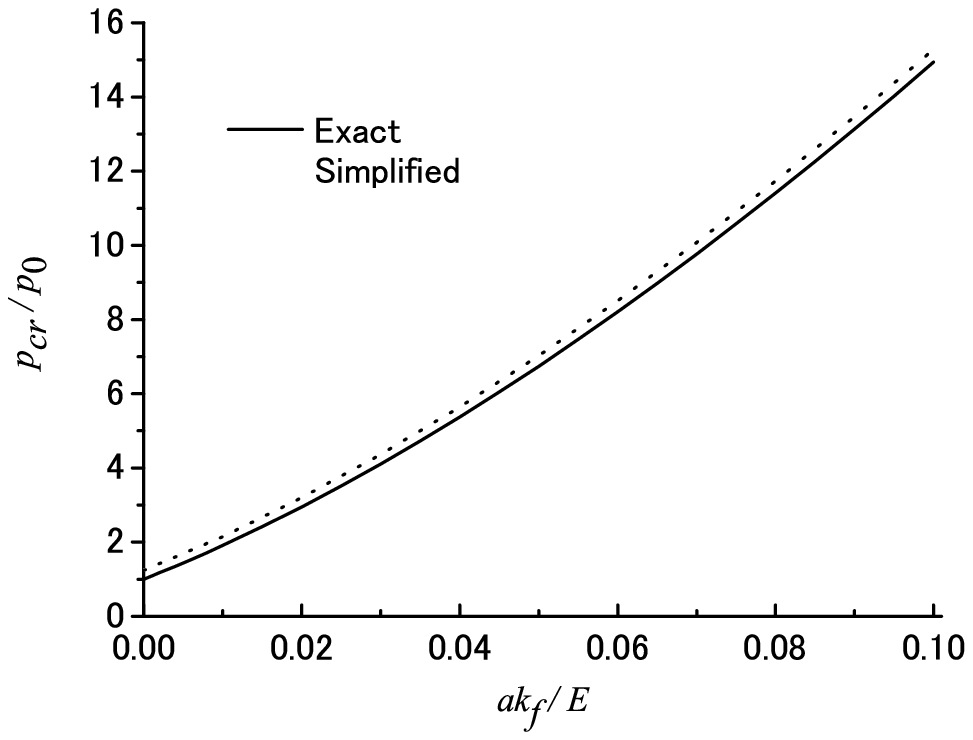}
\caption{Comparison of $p_{cr}$ ($a/h=100$).}
\label{Comparison(pcr)}
\end{figure}
\begin{figure}
\centering
\includegraphics[width=10.5cm]{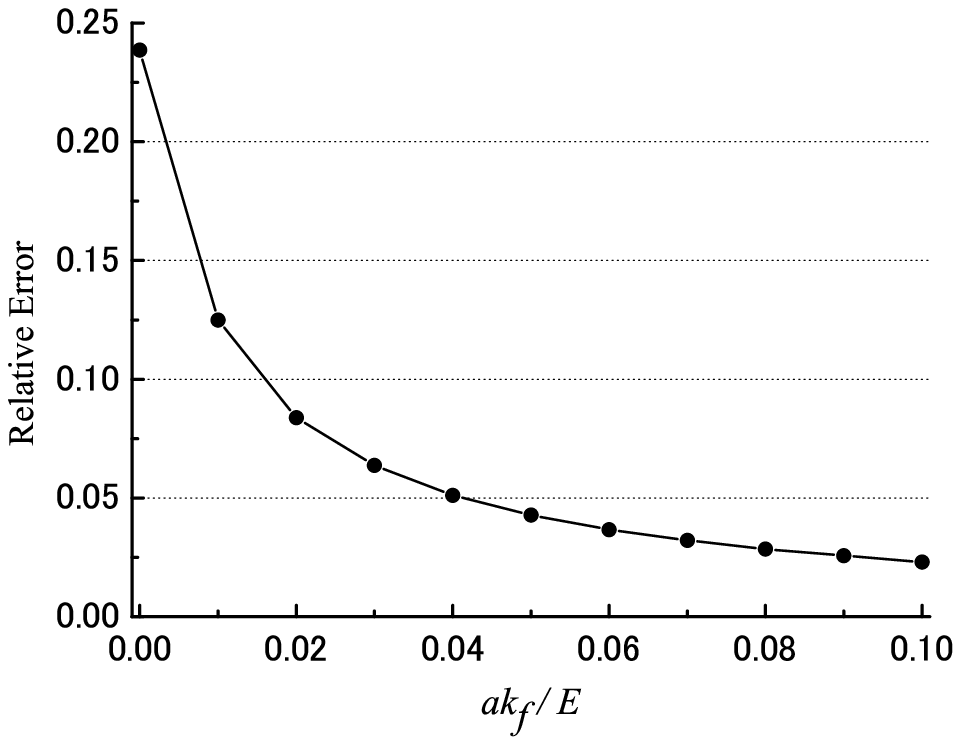}
\caption{Relative error of $p_{cr}$ ($a/h=100$).}
\label{Error(pcr)}
\end{figure}

\begin{figure}
\centering
\includegraphics[width=10.5cm]{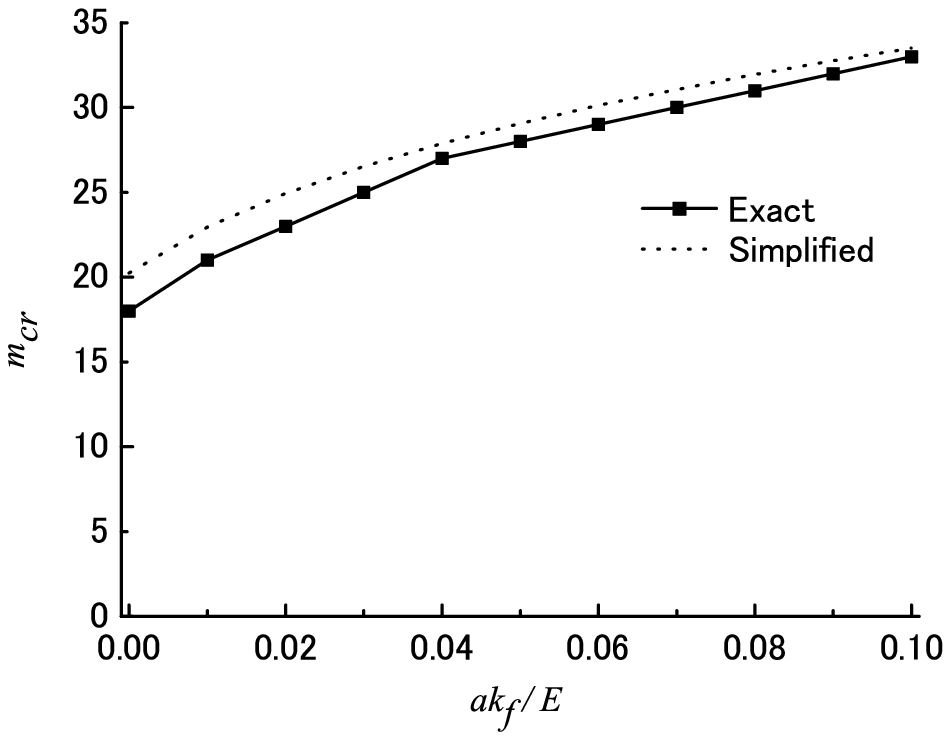}
\caption{Comparison of $m_{cr}$ ($a/h=100$).}
\label{Comparison(mcr)}
\end{figure}
\begin{figure}
\centering
\includegraphics[width=10.5cm]{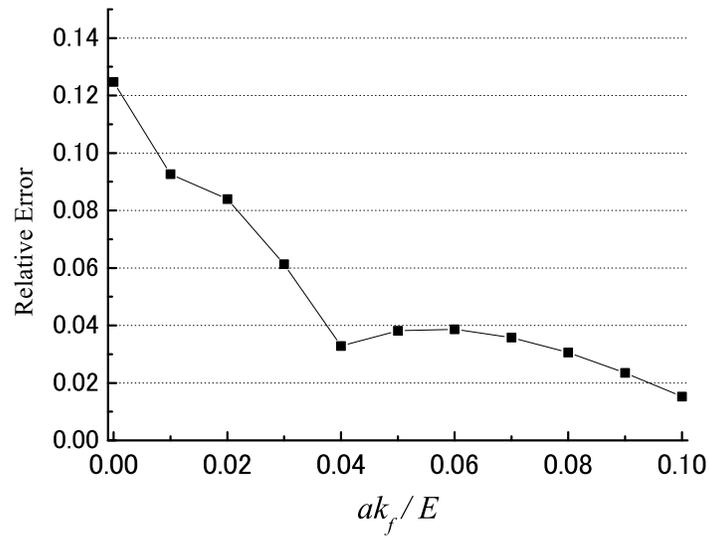}
\caption{Relative error of $m_{cr}$ ($a/h=100$).}
\label{Error(mcr)}
\end{figure}

Figure~\ref{Comparison(pcr)} shows the variation of the nondimensionalized
critical buckling pressure $p_{cr}/p_0$, in which $p_0$ is the
critical pressure for the empty complete spherical shell, which is
given by:
\begin{equation}
p_0 = \frac{2E}{\sqrt{3(1-\nu^2)}} \left(\frac{h}{a} \right)^2. 
\label{p_0}   
\end{equation}

The solid lines are the exact values obtained from Eq.(\ref{det2}) and
the dotted lines correspond to Eq.(\ref{pcr-s}) from the Rayleigh--Ritz
analysis.  As can be easily seen in this figure, the value of
$p_{cr}/p_0$ increases with increasing stiffness ratio $ak_f/E$ and
the exact and simplified $p_{cr}$ curve agree with each other.
Fig.~\ref{Error(pcr)} gives the relative error between exact and
simplified values, which is defined by
\begin{equation}
\frac{p_{cr(\mathrm{simplified})}-p_{cr(\mathrm{exact})}}{p_{cr(\mathrm{exact})}}.
\label{pcr-siplified}   
\end{equation}
The relative error converges with the increase of $ak_f/E$.

The comparison of the critical buckling mode number $m_{cr}$ obtained
from the exact and simplified approaches is shown in
Fig.~\ref{Comparison(mcr)}.  Larger values of $ak_f/E$ cause higher
buckling modes, as in the case of beams on an elastic Winkler
foudation \cite{WadeeHuntWhiting1997}.  The relative error
\begin{equation}
\frac{m_{cr(\mathrm{simplified})}-m_{cr(\mathrm{exact})}}{m_{cr(\mathrm{exact})}}.
\label{pcr-siplified}   
\end{equation}
is given in Fig.~\ref{Error(mcr)}.  We can also see the convergence
behaviour with the increase of $ak_f/E$.  These comparisons in
Figs.~\ref{Error(pcr)} and \ref{Error(mcr)} confirm the validity of
the proposed simplified formulations for $p_{cr}$ and $m_{cr}$
especially for higher stiffness ratios.

\subsection{Eigenmodes}

\begin{figure}
\centering
\includegraphics[width=10cm]{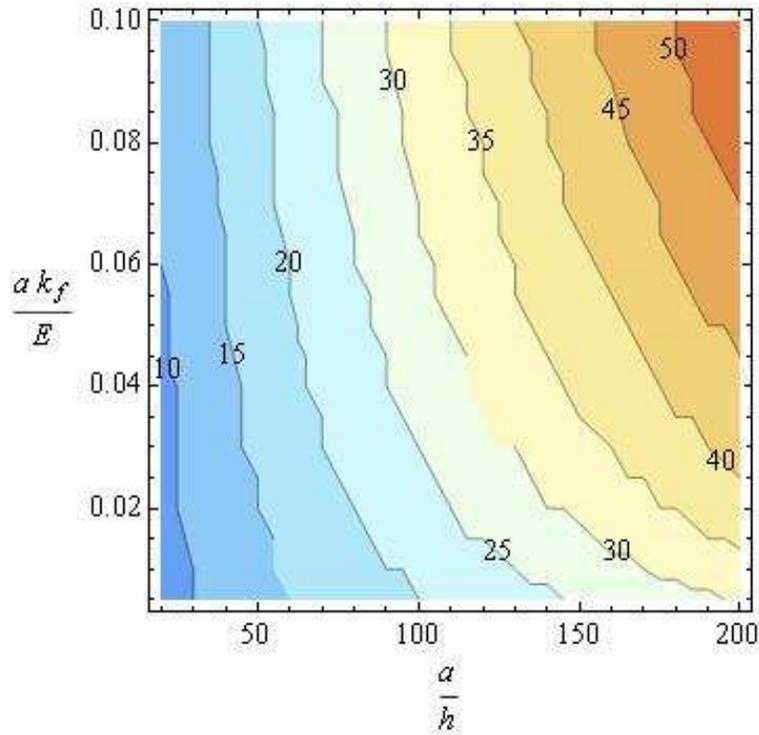}
\caption{Phase diagram of the buckling mode number $m$.  Various
  axisymmetric buckling modes can be found depending on the values of
  $ak_f/E$ and and $a/h$.}
\label{Phase}
\end{figure}

Figure \ref{Phase} shows a phase diagram of the buckling mode number
$m$.  This figure gives significant information on the various
axisymmetric buckling modes depending on the values of $ak_f/E$ and
and $a/h$.  It is observed from this figure that larger $ak_f/E$ and
$a/h$ values favour buckling modes with higher mode numbers.  We can
also find that when the spherical shells are relatively thin with
larger $a/h$ values, the effect of the foundation modulus becomes more
significant when compared with the case for the relatively thick shells
with smaller $a/h$ values.
It should be noted that The obtained mode number $m$=18 for 
the condition of $a/h=100$ and $ak_f/E=0.1$ is same as that found 
by Koiter \cite{KoiterPRDASB1969}.

\begin{figure}
\begin{tabular}{cc}
\begin{minipage}{0.5\hsize}
\centering
\includegraphics[width=5cm]{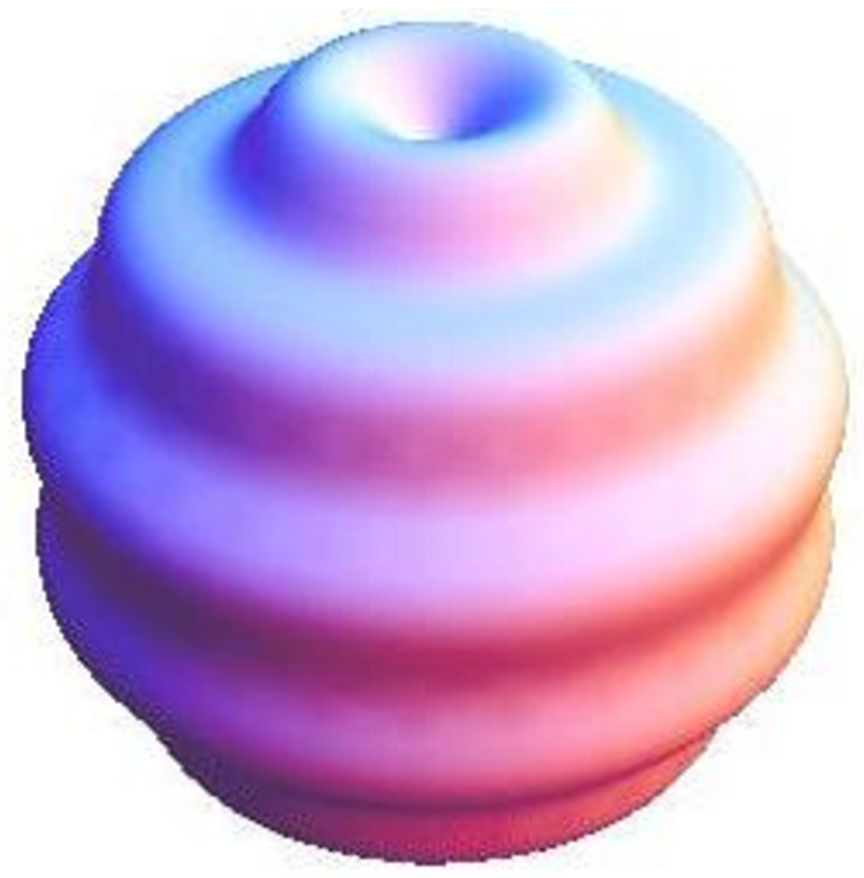}  
\end{minipage}
\begin{minipage}{0.5\hsize}
\centering
\includegraphics[width=4.2cm]{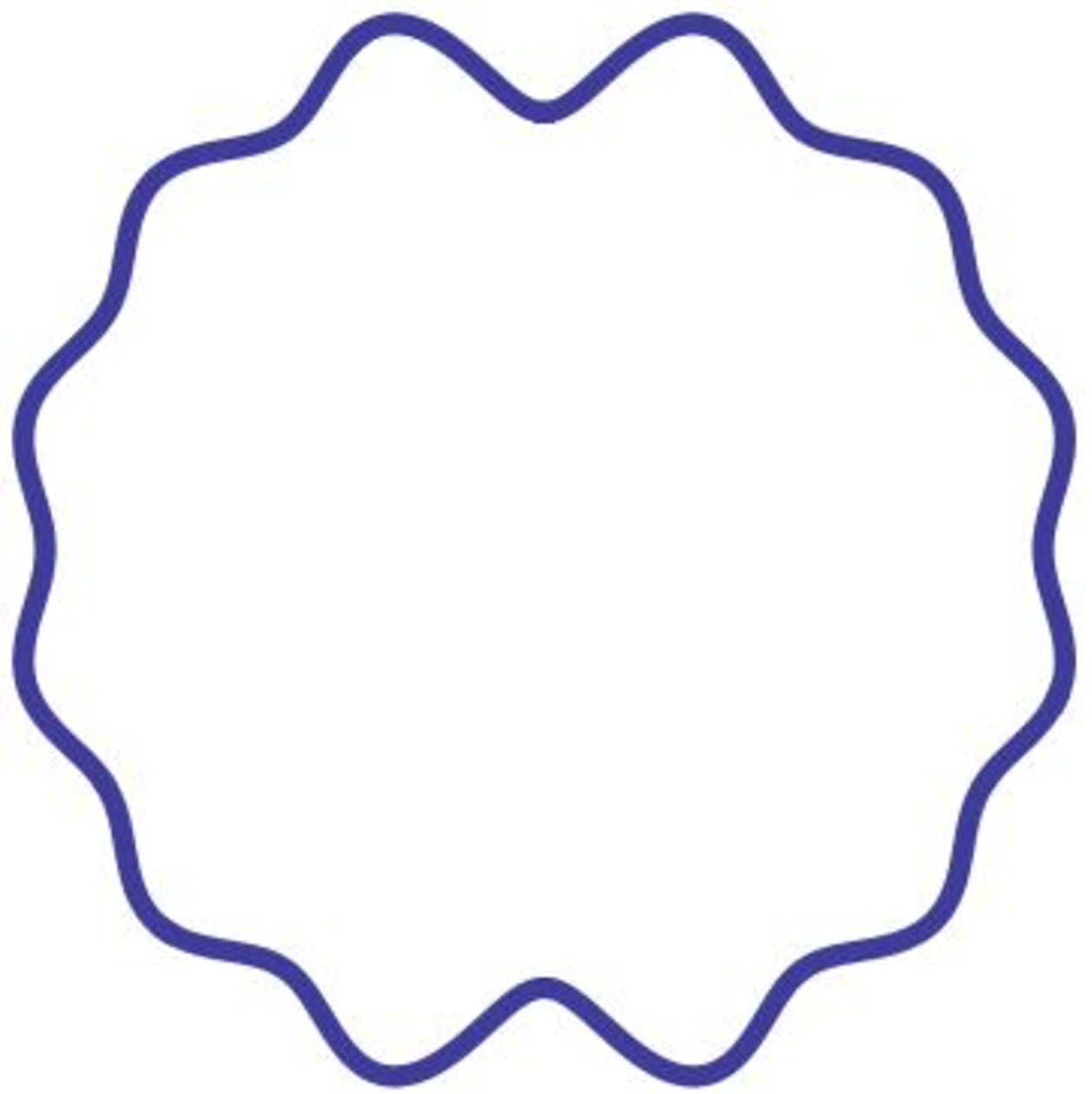}
\end{minipage}
\end{tabular}
\caption{Buckling modes of a hydrostatical pressurized "empty"
  complete spherical shell with $a/h=50$ and $ak_f/E=0$. (Left) 3-D
  view. (Right) Cross-sectional view through a great circle.  The
  buckling mode number $m = 12$ is found.}
\label{Sphere50-0}
\end{figure}

Figure \ref{Sphere50-0} illustrates the buckling eigenmodes for the empty
complete spherical shell of $a/h=50$.  The characteristic wavy-shaped
axisymmetric buckling deformation with the mode number $m=12$ can be
found in this case. 

\begin{figure}
\begin{tabular}{cc}
\begin{minipage}{0.5\hsize}
\centering
\includegraphics[width=5cm]{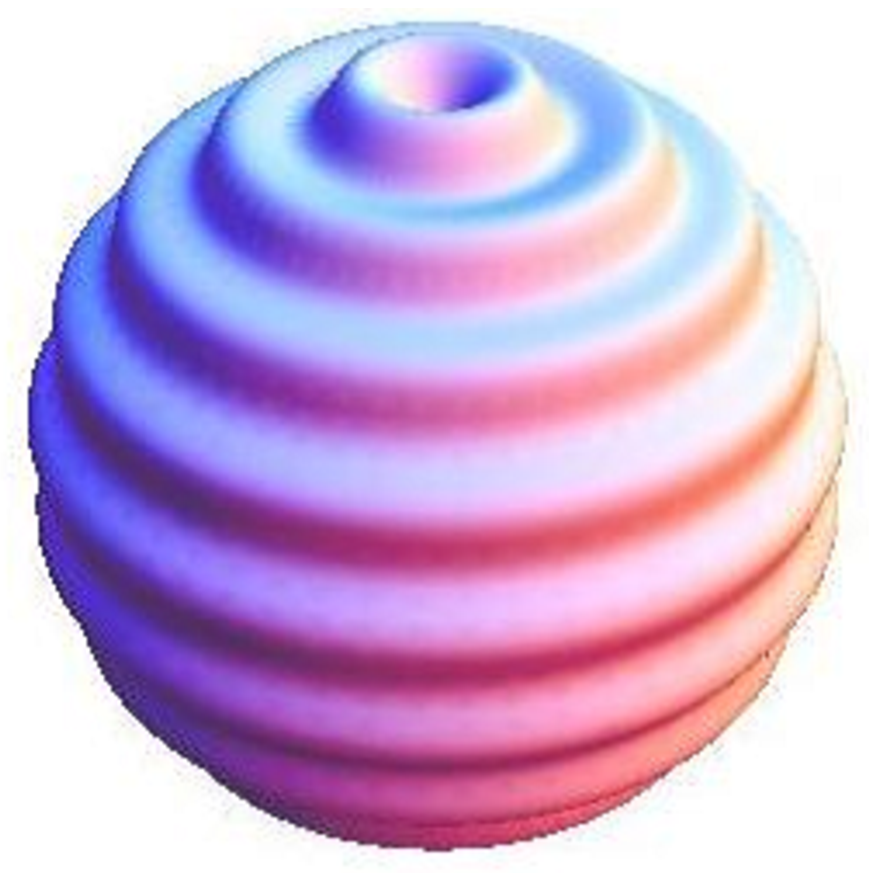}  
\end{minipage}
\begin{minipage}{0.5\hsize}
\centering
\includegraphics[width=4.2cm]{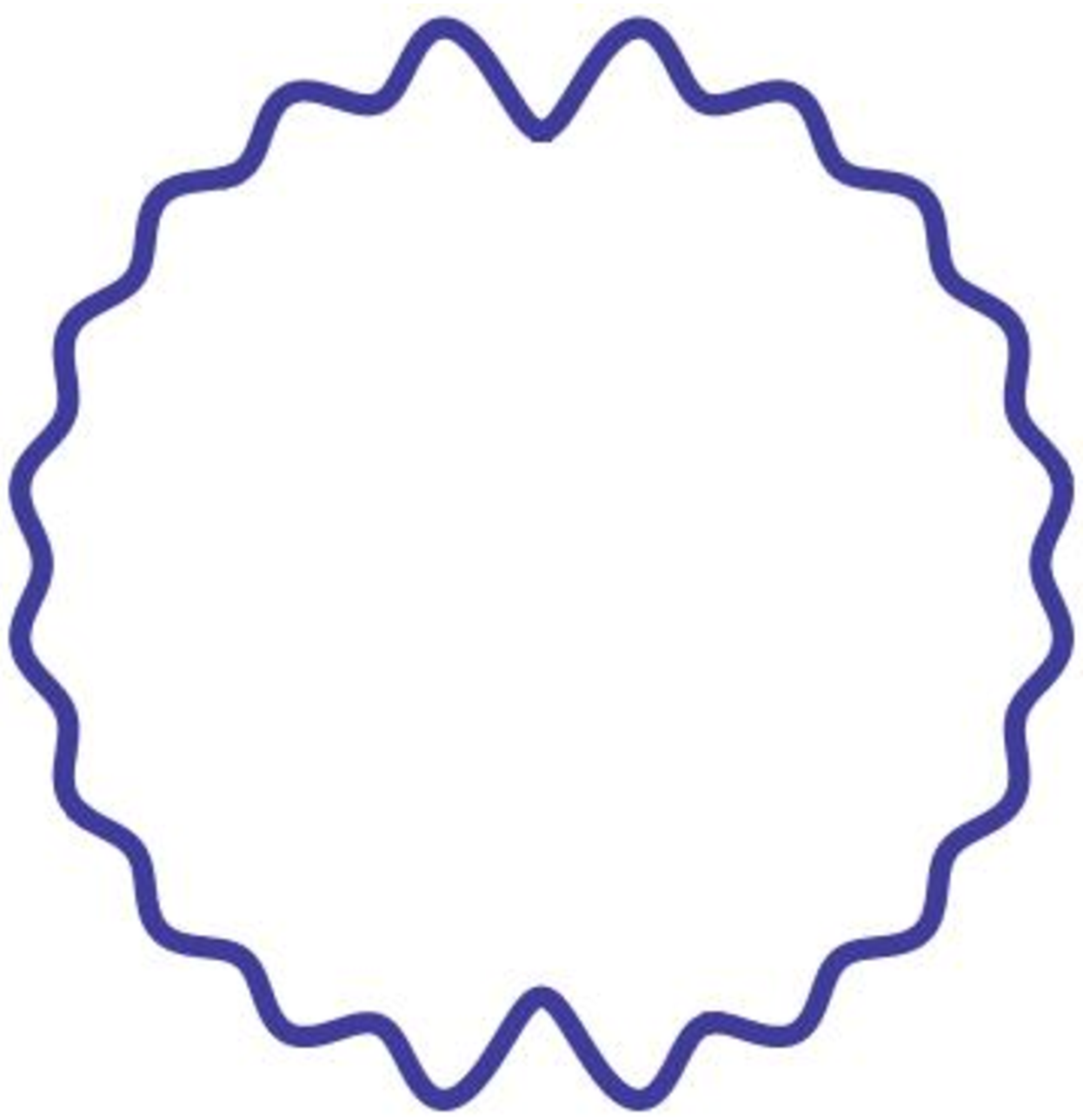}
\end{minipage}
\end{tabular}
\caption{Buckling modes of a hydrostatical pressurized "filled"
  complete spherical shell with $a/h=50$ and $ak_f/E=0.1$. (Left) 3-D
  view. (Right) Cross-sectional view through a great circle.  The
  buckling mode number $m = 20$ is found.}
\label{Sphere50-0.1}
\end{figure}

\begin{figure}
\begin{tabular}{cc}
\begin{minipage}{0.5\hsize}
\centering
\includegraphics[width=5cm]{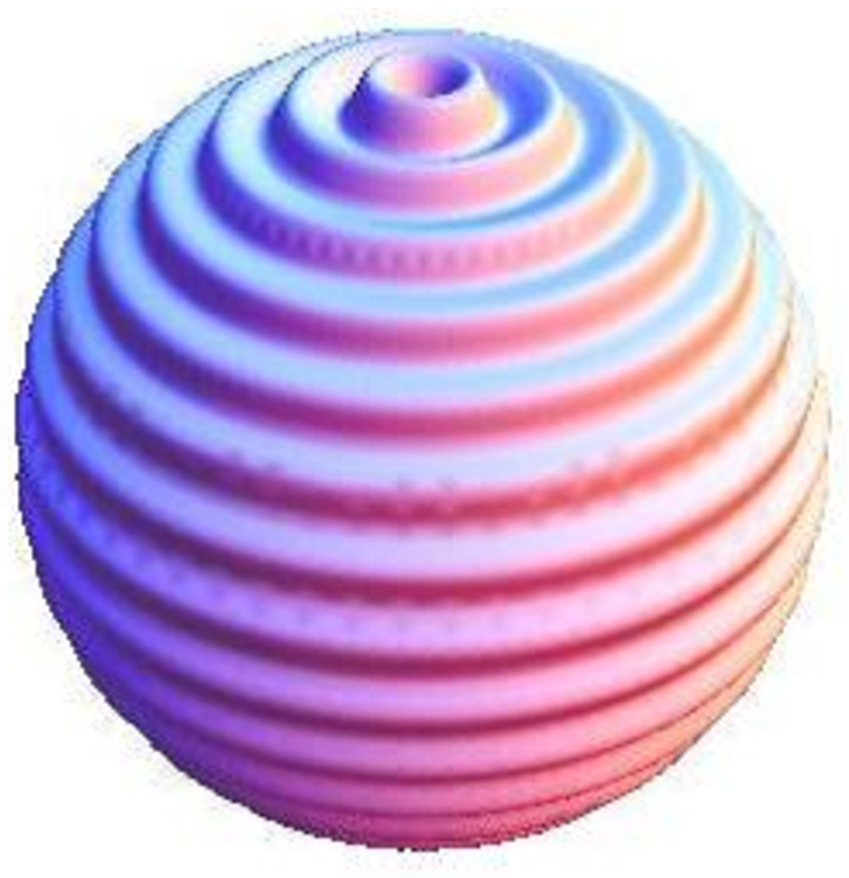}  
\end{minipage}
\begin{minipage}{0.5\hsize}
\centering
\includegraphics[width=4.2cm]{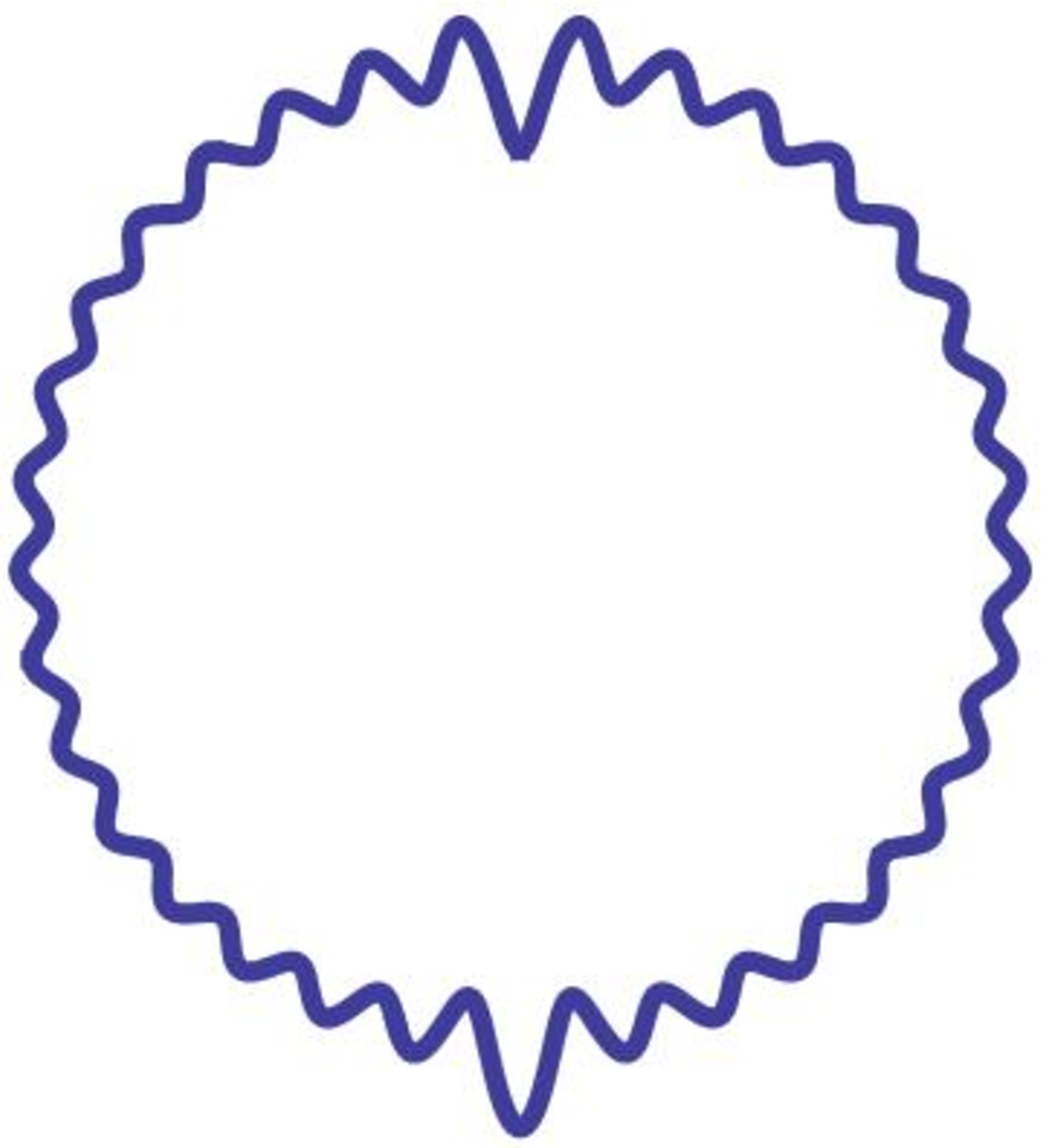}
\end{minipage}
\end{tabular}
\caption{Buckling modes of a hydrostatical pressurized "filled"
  complete spherical shell with $a/h=100$ and $ak_f/E=0.1$. (Left) 3-D
  view. (Right) Cross-sectional view through a great circle.  The
  buckling mode number $m = 33$ is found.}
\label{Sphere100-0.1}
\end{figure}

On the other hand, Figs.~\ref{Sphere50-0.1} and \ref{Sphere100-0.1}
show the eigenmodes for the ``filled'' spherical shell ($ak_f/E=0.1$)
with $a/h$=50 and 100, respectively; the axisymmetric buckling
deformation of the mode numbers $m=20$ and 33 are seen in these
figures.  As can be understood in the developed formulation,
non-axisymmetric modes cannot occur under the condition of constant
foundation modulus.  However, the cross-sectional shape for
Fig.~\ref{Sphere100-0.1} is moderately different from that for
Figs.~\ref{Sphere50-0} and \ref{Sphere50-0.1}.

The results shown in Figs.~\ref{Sphere50-0}--\ref{Sphere100-0.1}
suggest that the eigenmodes change corresponding to the mode number
$m$.  When $m$ is even, the corresponding deformation of the spherical
shell is symmetric about the equator, as shown in
Figs.~\ref{Sphere50-0} and \ref{Sphere50-0.1}.  On the other hand,
when $m$ is odd, the deformation is ``anti-symmetric'' about the
equator, that is, an inward deformation at the north pole is
accompanied by an outward deformation at the south pole.  This is due
to the property of the geometry of the sphere and the Legendre
functions, not whether the shell is empty or otherwise.


\section{Conclusions}

The characteristic buckling modes in hydrostatically pressurized
complete spherical shells filled with an elastic medium have been
demonstrated.  A theoretical formulation based on small-displacement
thin shell theory has derived governing equations of equilibrium that
have been solved using an exact methodology.  In addition, the
simplified formulations for estimating the critical pressure and the
mode number are proposed with a Rayleigh--Ritz approach.  The
formulations obtained from the Rayleigh--Ritz methodology is shown to
give sufficiently accurate results compared with the exact values.
    
From the formulation developed currently only axisymmetric eigenmodes
can occur in spite of adding the Winkler foundation term; critical
modes that are symmetric or anti-symmetric about the equator may be
determined depending on the combination of the foundation stiffness
and the radius to wall-thickness ratios. A phase diagram has been
established to obtain the requisite values of $a/h$ and $ak_f/E$ for
observing a corresponding buckling mode.

This work is a fundamental investigation that has been developed with
future studies in mind. These will address the post-buckling behaviour
\cite{Thompson1973,Thompson1984} and non-axisymmetic modes which may
occur when a foundation modulus is not a constant value and a function
of the wave number of the shell deformation. The Rayleigh--Ritz approach
developed currently is more likely to be employed for the nonlinear
studies given that it allows for a simpler solution methodology;
it has been shown in the present work that it provides an
excellent approximation to the linear case. This will also provide
an insight into the electrical properties of carbon onions,
{\it i.e.,} quasi-spherical nanoparticles consisting of
concentric graphitic shells.  They possess a high specific surface
area ($\sim$ 500 m$^2$ per gram) with no porosity and high
field-emission stability owing to the endurance of carbon-carbon
atomic bonds.  These two features are appealing in the design of
electrochemical capacitors \cite{PechNatureNTN2010} and point electron
sources \cite{MSWangACSNano2010}, in which the radial buckling of
outer layers may lead positive effects in operation.  It was also
suggested in \cite{PudlakPRA2009} that the electronic structure of
carbon onions is sensitive to the radii of concentric layers and their
separations.  Hence, their asymmetric buckling under pressure should
cause a significant change in the electronic states, where the
mechanical deformation of the underlying carbon structure induces a
effective electrostatic potential that exerts on the mobile electrons
on the layers \cite{ShimaPRB2010g,OnoPRB2010}.  The present findings
will be helpful for addressing the deformation-driven carbon-onion
behaviours from both academic and practical viewpoints.  

\section*{Acknowledgements}

We acknowledge the contribution of Professors Takashi Mikami and
Shunji Kanie of Hokkaido University for stimulating discussions and
helpful comments.  This study was conducted during the 6 months that
the leading author (MS) spent within the Structures Section of the
Department of Civil and Environmental Engineering at Imperial College
London as a Visiting Research Fellow.  Financial support from Faculty
of Engineering, Hokkaido University for the stay is greatly
acknowledged.  
The last author (HS) thanks the financial supports from the Sumitomo Foundation
and the Suhara Memorial Foundation.
This study was also supported by a Grant-in-Aid for
Scientific Research from MEXT, Japan.

\appendix
\section{The coefficients of the governing equations}

The coefficients of the governing equations 
(\ref{ODEu-direction})--(\ref{ODEw-direction}) are written as follows:

\begin{align}
  & a_u^{(0)} = (1+k)\left[\frac{1-\nu}{2}(1-\cot^2\phi)\sin\phi -
  \frac{n^2}{\sin\phi} \right],    \nonumber \\ 
  & a_u^{(1)} = (1+k)\frac{1-\nu}{2} \cos\phi,  \
  a_u^{(2)} = (1+k)\frac{1-\nu}{2} \sin\phi,  \  
  a_u^{(3)} = a_u^{(4)} = 0,      \nonumber \\
  & b_u^{(0)} = (1+k)\frac{\nu-3}{2} n \cot\phi, \  
  b_u^{(1)} = -(1+k)\frac{1+\nu}{2} n,    \
  b_u^{(2)} = b_u^{(3)} = b_u^{(4)} = 0,   \nonumber \\
  & c_u^{(0)} = k n \left( 2 - \frac{n^2}{\sin^2\phi} \right),    \ 
  c_u^{(1)} = k n \cot\phi, \     
  c_u^{(2)} = k n,             \
  c_u^{(3)} = c_u^{(4)} = 0,      \nonumber \\
  & a_v^{(0)} = (1+k)\frac{\nu-3}{2} n \cot\phi, \   
  a_v^{(1)} = (1+k)\frac{1+\nu}{2} n,    \
  a_v^{(2)} = a_v^{(3)} = a_v^{(4)} = 0,   \nonumber \\
  & b_v^{(0)} = -\frac{1+k}{\sin\phi} \left( \cos^2\phi +
    \nu\sin^2\phi + \frac{1-\nu}{2}n^2 \right),   \nonumber \\ 
  & b_v^{(1)} = (1+k)\cos\phi, \     
  b_v^{(2)} = (1+k)\sin\phi,      \
  b_v^{(3)} = b_v^{(4)} = 0,      \nonumber \\
  & c_v^{(0)} = -2 k n^2 \frac{\cot\phi}{\sin\phi},  \
  c_v^{(1)} = -k \left( 2 - \frac{1+n^2}{\sin^2\phi} \right)\sin\phi,
  \nonumber \\ 
  & c_v^{(2)} = -k \cos\phi,  \
  c_v^{(3)} = -k \sin\phi,  \ 
  c_v^{(4)} = 0,    \nonumber \\
  & a_w^{(0)} = (1+k)(1+\nu)n - k n \left( 2 +
    \frac{1-n^2}{\sin^2\phi} \right),    \nonumber \\
  & a_w^{(1)} = k n \cot\phi, \
  a_w^{(2)} = - k n,  \
  a_w^{(3)} = a_w^{(4)} = 0,      \nonumber \\
  & b_w^{(0)} = (1+k)(1+\nu)\cos\phi - k\cos\phi \left( 2 +
    \frac{1-n^2}{\sin^2\phi} \right),   \nonumber \\ 
  & b_w^{(1)} = (1+k)(1+\nu)\sin\phi + \frac{k (n^2 +
    \cos^2\phi)}{\sin\phi},    \nonumber \\ 
  & b_w^{(2)} = -2k \cos\phi,   \
  b_w^{(3)} = - k \sin\phi,   \
  b_w^{(4)} = 0,    \nonumber \\
  & c_w^{(0)} = -k \Bigg[2(1+\nu)\sin\phi +
  n^2\frac{3-\nu+4\cot^2\phi}{\sin\phi}  
  - \frac{n^4}{\sin^3\phi} \Bigg],  \nonumber \\
  & c_w^{(1)} = \left( 1 - \nu + \frac{1+2n^2}{\sin^2\phi} \right)
  \cos\phi,   \ 
  c_w^{(2)} = -\frac{1 + 2n^2 + \nu \sin^2\phi}{\sin\phi},  \nonumber \\ 
  & c_w^{(3)} = 2k \cos\phi, \ c_w^{(4)} = k \sin\phi.       
\label{coefficients}   
\end{align}





\bibliographystyle{model1a-num-names}
\bibliography{<your-bib-database>}




\end{document}